\documentclass{elsart}
\usepackage[usenames,dvips]{pstcol}
\usepackage{color}
\usepackage{amssymb}
\usepackage{epsfig}
\usepackage{footnote}
\usepackage{longtable}

\newrgbcolor{maroon}{.45 0.1 0.1}
\begin{document}
\begin{frontmatter}
\title{Signal Processing for Pico-second Resolution Timing Measurements}
\author[UC]{Jean-Francois Genat}
\author[Hawaii]{Gary Varner}
\author[UC]{Fukun Tang}
\author[UC]{Henry Frisch}
\address[UC]{Enrico Fermi Institute, University of Chicago\\
5640 S. Ellis Ave, Chicago IL, 60637}
\address[Hawaii]{University of Hawaii, 2505 Correa Road, Honolulu, HI, 96822}

\begin{abstract}
    The development of large-area homogeneous photo-detectors with
    sub-millimeter path lengths for direct Cherenkov light and for
    secondary-electrons opens the possibility of large
    time-of-flight systems for relativistic particles with
    resolutions in the pico-second range.  Modern ASIC techniques
    allow fast multi-channel front-end electronics capable of
    sub-pico-second resolution directly integrated with the
    photo-detectors. However, achieving resolution in the
    pico-second range requires a precise knowledge of the signal
    generation process in order to understand the pulse waveform,
    the signal dynamics, and the noise induced by the detector
    itself, as well as the noise added by the processing
    electronics. Using the parameters measured for fast
    photo-detectors such as micro-channel plates
    photo-multipliers, we have simulated and compared the
    time-resolutions for four signal processing techniques:
    leading edge discriminators, constant fraction discriminators,
    multiple-threshold discriminators and pulse waveform sampling.
\end{abstract}
\end{frontmatter}

\section{Introduction}
The typical resolution for measuring time-of-flight of relativistic
particles achieved in large detector systems in high energy physics
has not changed in many decades, being on the order of 100
psec~\cite{CDF,TOF_review}. This is set by the characteristic scale
size of the light collection paths in the system and the size of the
drift paths of secondary electrons in the photo-detector itself, which
in turn are usually set by the transverse size of the detectors,
characteristically on the order of one inch (100 psec).  However, a
system built on the principle of Cherenkov radiation directly
illuminating a photo-cathode followed by a photo-electron amplifying
system such as a Micro-Channel Plate Photo-Multiplier
(MCP-PMTs)~\cite{Wiza} with characteristic dimensions of 10 microns or
less, has a much smaller characteristic size, and consequently a much
better intrinsic time resolution~\cite{Credo_Rome,Inami,Vavra}.

\begin{figure}[!t]
\centering
\includegraphics[angle=0,width=0.65\textwidth]
{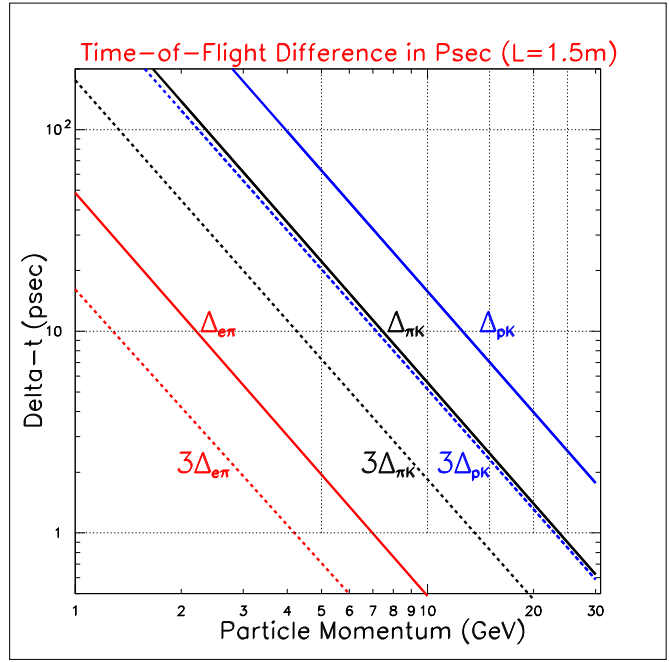}
\caption{ The difference in times (solid lines) over a path-length of
  1.5m for electrons (i.e. zero time delay relative to the speed of
  light) versus pions, pions versus kaons, and kaons versus protons,
  as a function of the charged particle momentum. The time resolutions
  necessary for a 3-$\sigma$ separation versus momentum are also shown
  (dashed lines).}
\label{fig:fig0_separability}
\end{figure}

Time-of-flight techniques with resolution of less than several
picoseconds would allow the measurement of the mass, and hence the
quark content, of relativistic particles at upgraded detectors at high
energy colliders such as the Fermilab Tevatron, the LHC, Super-B
factories, and future lepton-colliders such as the ILC or a
muon-collider, and the association of a photon with its production
vertex in a high-luminosity collider~\cite{LHC_photon}. Other new
capabilities at colliders would be in associating charged particles
and photons with separate vertices in the 2-dimensional
time-vs-position plane, and searching for new heavy particles with
short lifetimes~\cite{gunion,toback_delayed_photon}.  The difference
in transit times over a path-length of 1.5 meters, typical of the
transverse dimension in a
solenoidal collider detector such as CDF or ATLAS, is shown in
Figure~\ref{fig:fig0_separability}. Many other applications with
different geometries, such as forward spectrometers, would have
significantly longer path-lengths, with a consequent reach in
separation to higher momenta, as can be scaled from the figure.

 There are possible near-term applications of fast timing requiring
 resolutions of several picoseconds but smaller area systems
 ($\simeq$0.001-1 m$^2$), such as missing-mass searches for the
 Higgs at the LHC~\cite{LHC_forward}, and non-magnetic spectrometers for
 the development of 6-dimensional phase-space muon
 cooling~\cite{MANX}.  There are likely to be applications in other
 fields as well, such as measuring longitudinal emittances in
 accelerators, precision time-of-flight in mass spectroscopy in
 chemistry and geophysics, and applications in medical imaging.

At lower time and position resolution, the same techniques could be
 used for instrumenting the surfaces of large-ring imaging water
 Cherenkov counters, in which measurement of both the position and
 time-of-arrival of Cherenkov photons would allow reconstruction
 of track directions and possibly momenta~\cite{nicholson}.

In order to take advantage of photo-detectors with intrinsic single
photo-electron resolutions of tens of picoseconds to build large-area
time-of-flight systems, one has to solve the problem of collecting
signal over distances large compared to the time resolution while
preserving the fast time resolution inherent in the small feature size
of the detectors themselves.  Since some of these applications would
cover tens of square meters and require tens of thousands of detector
channels, the readout electronics have to be integrated via
transmission lines with the photo-detector itself in order to reduce
the physical dimensions and power, increase the analog band-width,
improve readout speed, and provide all-digital data output.

There are a number of techniques to measure the arrival time of very
fast electrical pulses~\cite{porat,genat,kalisz,mantyniemi}.
Typically one measures the time at which the pulse crosses a single
threshold, or, for better resolution, the time at which the pulse
reaches a constant fraction of its amplitude~\cite{cova}. An extension
of the threshold method is to measure the time that a pulse crosses
multiple thresholds~\cite{heejong}.

 A recent development is the large-scale implementation of fast analog
 waveform sampling onto arrays of storage capacitors using CMOS
 integrated circuits at rates on the order of a few GSa/s. Most, if not all of them, 
have actually 3dB analog bandwidths below 1 GHz
~\cite{breton,delagnes,ritt,varner}.
 The steady decrease in feature size and power for custom integrated
 circuits now opens the possibility for multi-channel chips with
 multi-GHz analog bandwidths, and able to
 sample between 10 and 100 GHz, providing both time and amplitude
 after processing. Assuming that the signals are recorded over a time
 interval from before the pulse to after the peak of the pulse, with
 sufficient samples fast waveform sampling provides the information to
 get the time of arrival of the first photo-electrons, the shape of
 the leading edge, and the amplitude and integrated charge.  While
 other techniques can give time, amplitude, or integrated charge, fast
 sampling has the advantage that it collects all the information, and
 so can support corrections for pileup, baseline shifts before the
 pulse, and filtering for noisy or misshapen pulses. In applications
 such as using time-of-flight to search for rare slow-moving
 particles, having the complete pulse shape provides an important
 check that rare late pulses are consistent with the expected waveform.

The outline of this note is as follows: 
Section~\ref{timing_techniques} describes the four techniques for
determining the time-of-arrival of an electrical pulse from a photo-detector.
Section~\ref{Simulations} describes
the input signal parameters of the simulation program used for the comparisons, 
and the parameters used for each
of the four methods in turn. Section~\ref{results} presents the results 
and the methods and parameters to be used
in real systems. The conclusions and summary are given in 
Section~\ref{conclusions}.

\section{Timing  techniques}
\label{timing_techniques} Present photo-detectors such as micro-channel 
plate photo-multipliers (MCP-PMTs) and silicon
photomultipliers achieve rise-times well below one 
nanosecond~\cite{photek,inami,bondarenko}. Ideal timing readout
electronics would extract the time-of-arrival of the first charge 
collected, adding nothing to the intrinsic
detector resolution. Traditionally the best ultimate performance 
in terms of timing resolution has been obtained
using constant fraction discriminators (CFDs) followed by high 
precision amplitude digitization. However, these
discriminators make use of wide-band delay lines that cannot be 
integrated easily into silicon integrated
circuits, and so large  front-end readout systems using CFD's to 
achieve sub-nsec resolution have are not yet been
implemented.

Several other well-known techniques in addition to constant-fraction
discrimination have long been used for timing
extraction of the time-of-arrival of a pulse:

\begin{enumerate}
\setlength{\itemsep}{-0.03in}
\item Single threshold on the leading edge;
\item Multiple thresholds on the leading edge, followed by a
fit to the edge shape;
\item Pulse waveform sampling, digitization and pulse reconstruction.
\end{enumerate}

Applying a fixed threshold to the leading edge, which is a
one-parameter technique, suffers from a dependence of the extracted
time with the pulse amplitude, even for identical waveforms.  
In addition, this method is sensitive to base-line shifts due to
‘pile-up’, the overlap of a pulse with a preceding one or many, a
situation common in high-rate environments such as in collider
applications. Also, for applications in which one is searching for
rare events with anomalous times, the single measured time does not give
indications of possible anomalous pulse shapes due to intermittent
noise, rare environmental artifacts, and other real but rare
annoyances common in real experiments.

In contrast, constant fraction discrimination takes into account
the pulse amplitude. The most commonly used constant fraction
discriminator technique forms  the difference between attenuated and
delayed versions of the original signal. There are therefore three
parameters: the delay, the attenuation ratio, and the threshold. These
parameters have to be carefully set with respect to the pulse
characteristics in order to obtain the best timing resolution.

The multiple threshold technique samples the leading edge at
amplitudes set to several values, for instance at values equally
spaced between a minimum and a maximum threshold. The leading edge is
then reconstructed from a fit to the times the pulse reaches the
thresholds to extract a single time as characteristic of the pulse. As
in the case of constant fraction discrimination, if the pulse shape is
independent of amplitude, the reconstructed time will also not depend
upon the pulse amplitude, provided the thresholds are properly set.\\
In the simulation, the single threshold was set at 8$\%$ of the average
pulse amplitude, providing the best timing resolution. Lower threshold 
values could not be used due to the noise, particularly at low photo-electrons 
numbers. For multiple-threshold, almost no improvement was found at more
than four thresholds; the lowest and highest thresholds were determined
to avoid sensitivity and inefficiencies at low photo-electrons numbers.

Waveform sampling stores successive values of the pulse waveform. 
For precision time-of-arrival measurements, such
as considered here, one needs to fully sample at least the leading 
edge over the peak. In order to fulfill the
Shannon-Nyquist condition ~\cite{shannon}, the sampling period has 
to be chosen short enough to take into account all frequency
components containing timing information, which is that the minimum 
sampling frequency is set at least at twice the highest
frequency in the signal's Fourier spectrum. In practice, there are frequency 
components contributing to the leading edge
well above the 3-dB bandwidth of the signal spectrum, before the noise 
is dominant, and these components should not be filtered out. After
digitization, using the knowledge of the average waveform, pulse 
reconstruction allows reconstructing the edge or
the full pulse with good fidelity. The sampling method is unique 
among the four methods in providing the pulse
amplitude, the integrated charge, and figures of merit on the 
pulse-shape and baseline, important for detecting
pile-up or spurious pulses.

\section{Simulations}
\label{Simulations}

We have developed a Monte-Carlo simulation tool using
MATLAB~\cite{matlab} in order to generate pulses having the temporal
and spectral properties of fast photo-detector signals, and to
simulate and compare the behavior of the four techniques described in
Section II.  Both amplitude and timing resolution are estimated as a
function of various parameters, such as the number of photo-electrons,
the signal-to-noise ratio, and the analog bandwidth of the input
section of the front-end electronics~\cite{definitions}. In the case
of sampling, the resolution is estimated as a function of the sampling
frequency, the number of bits in the analog-to-digital conversion, and
the timing jitter of the sampling.  The four methods are simulated and
results are evaluated with respect to each other below.

\begin{figure}[!t]
\centering
\includegraphics[angle=0,width=0.65\textwidth]
{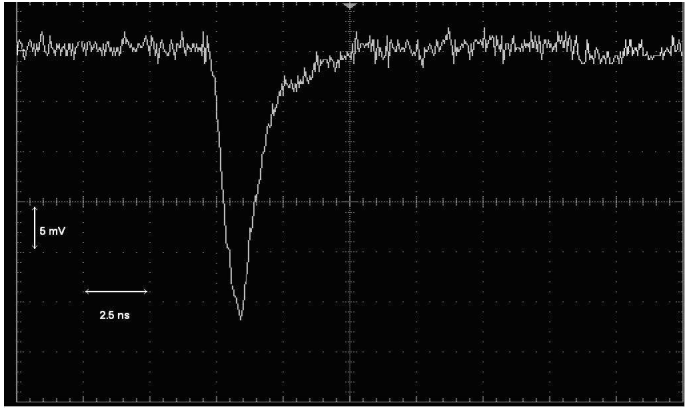}
\caption{ A signal from a Photonis XP85011 micro-channel plate
  photo-detector with $25\mu$-diameter pores~\cite{Photonis}, recorded
  with a Tektronix TDS6154C oscilloscope~\cite{Tektronix}, using the
  ANL laser test-stand~\cite{ANL_laser}.  Signal to Noise ratio is 40. The oscilloscope analog
  bandwidth is 15 GHz, sampling rate 40 GS/s, and the horizontal and vertical scales are 2.5
  ns/division, and 5mV/division, respectively.}
\label{fig:fig1_scope_trace}
\end{figure}

\subsection{Input signals}
In order to run a Monte-Carlo using realistic signals, pulses have
been synthesized based on measurements of MCP signals such as
those shown in Figure~\ref{fig:fig1_scope_trace}. The synthesized
signals are calculated as the convolution of a triangular waveform
having a rise time of 100 ps and fall time on the same order~\cite{photek},
with an impulse waveform of $\tau e^{-t/\tau}$, where $\tau$ is set
according to the analog bandwidth of the front-end electronics.
With a 1.5 GHz analog bandwidth, $\tau$ is set to 235 ps.
This waveform is then convolved with itself, in order to match
the MCP pulse shape. The simulated input
signals have variable spread in amplitude, implemented by taking into
account the number of incident photo-electrons, ${N_{pe}}$, as
normally distributed with $\sigma$ proportional to $1/\sqrt{N_{pe}}$.
To average over discrete binning effects, we introduce a spread in the
initial time, distributed uniformly between +$\tau$ and –$\tau$.

The simulation includes both detector shot noise and thermal noise
superimposed on the signal. The noise is taken with two
contributions:
\begin{enumerate}
\setlength{\itemsep}{-0.03in}
\item white shot noise from the MCP, which is then shaped by the
electronics in the same way as is the MCP output signal;
\item white thermal noise is assumed to originate from the
electronics components.
\end{enumerate}

\begin{figure}[!th]
\centering
\includegraphics[angle=0,width=0.65\textwidth]{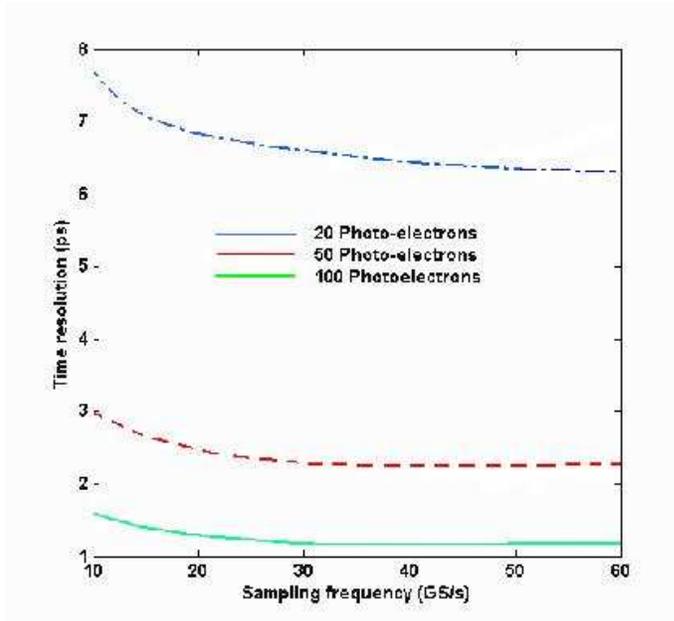}
\caption{Typical synthesized MCP-PMTs signals used in the simulation for an
input signal of 50 photo-electrons. To average over discrete binning effects, 
signals are spread in the
initial time, distributed uniformly between +$\tau$ and –$\tau$.}
\label{fig:fig2_300_traces}
\end{figure}

These two noise spectra are weighted so that they contribute
equally to the overall signal-to-noise ratio.

\begin{figure}[!bh]
\centering
\includegraphics[angle=0,width=0.65\textwidth]{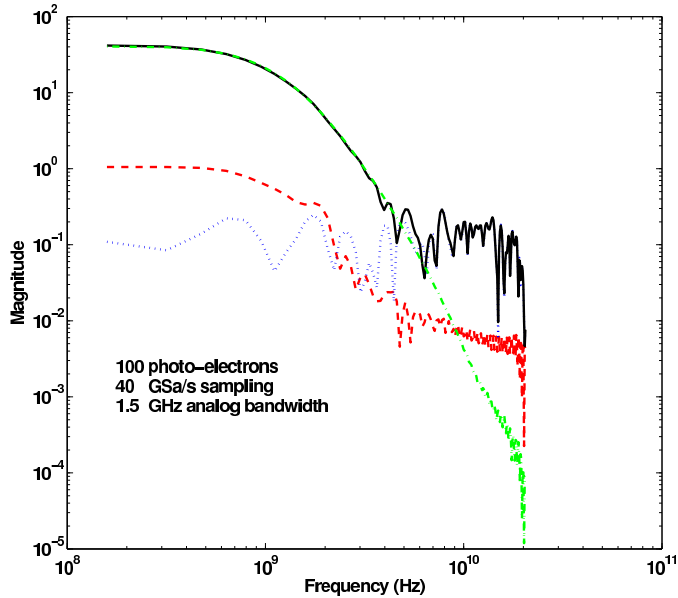}
\caption{Fourier spectra of a synthesized noiseless MCP signal
(green, dashed-dots), MCP shot noise (red, dashed), electronics noise (blue, dots), and the final
`noisy' MCP signal with both sources of noise included (black, solid).
Sampling rate is set at 40 GSa/s, therefore, all frequency components
above 20 GHz have been removed with an anti-aliasing filter.
}
\label{fig:fig3_noise_spectra}
\end{figure}

A set of 300 synthesized signals with a mean of 50 photo-electrons,
assuming a 1.5 GHz analog bandwidth, is shown in
Figure~\ref{fig:fig2_300_traces}. Figure~\ref{fig:fig3_noise_spectra}
displays the Fourier spectra of: a)a noiseless MCP signal, b) MCP shot
noise, c) electronics noise, and d) the final noisy MCP signal,
including both sources of noise, for 20 input photo-electrons, and an
overall bandwidth of 1.5 GHz. The signal-to-noise ratio is taken to be
32.

\subsection{Simulation of the Leading Edge Discriminator}
In the simulation the fully-simulated waveform (i.e. with noise added
as described above) is intersected with a threshold set at the value
providing the best timing resolution for a given set of external
parameters, such as the analog bandwidth, or the number of
photo-electrons. For comparison with actual threshold measurements, some overdrive effects
should probably be considered. The comparators are assumed to be ideal, i.e.  the
trigger is generated by the first time step in the simulation that has
a pulse amplitude over the `arming' threshold. The threshold is set
between 4\% and 15\% of the average amplitude pulse height.

\subsection{Simulation of the Multi-Threshold Discriminator}

\begin{figure}[!t]
\centering
\includegraphics[angle=0,width=0.65\textwidth]{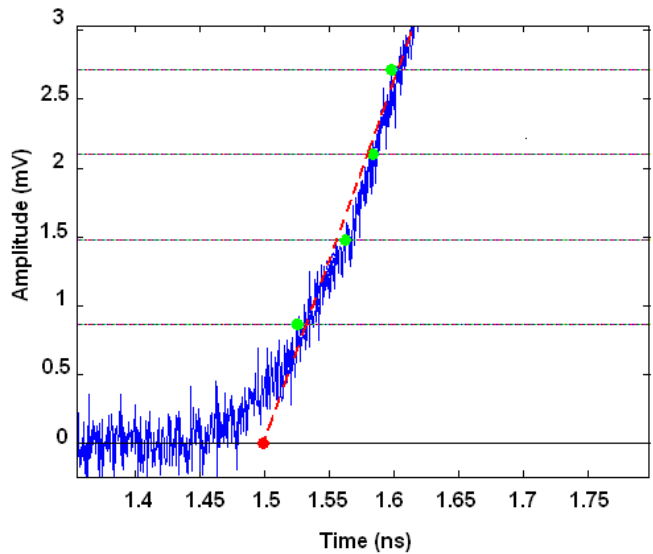}
\caption{ An example of the multi-threshold simulation and
technique. Green dots: crossed thresholds, red dots: zero-crossing time
from the extrapolation of the fit (red line) to the nominal start of
the pulse at zero amplitude.  }
\label{fig:fig4_multi_threshold}
\end{figure}

The multiple threshold technique~\cite{heejong} intersects the input waveform
with several thresholds, typically (but not necessarily)
equally spaced between a minimum and a maximum. A fit is then
performed, and a single time is extracted as the time-of-arrival of
the pulse. Figure~\ref{fig:fig4_multi_threshold} shows an example,
with the best-fit time taken as the intersection of the fit
extrapolated to the intersection with the time axis.  For the input
parameters we have chosen, we find that four thresholds, equally
spaced between 10 and 50\% of the average pulse height, are enough so
that more thresholds do not significantly improve the performance.

\subsection{Simulation of the Constant Fraction Discriminator}
A constant fraction discriminator fires at a fixed fraction of the
amplitude of the pulse, relying on the assumption that the pulse shape
is independent of amplitude. The implementation considered here is that the
input pulse is attenuated between 30 and 40\%, inverted, and then
summed with a version of the pulse delayed between 150 and 200
picoseconds.  If the pulse passes a predetermined `arming' threshold, set in
the simulations to between 10 and 20\% of the average pulse height,
the time of the zero-crossing of the summed pulse is measured.  As in
the case of the leading edge simulation, the parameters are optimized
to get the best possible timing resolution for a given set of
parameters.

\subsection{Simulation of Pulse Waveform Sampling}

\begin{figure}[!b]
\centering
\includegraphics[angle=0,width=0.45\textwidth]{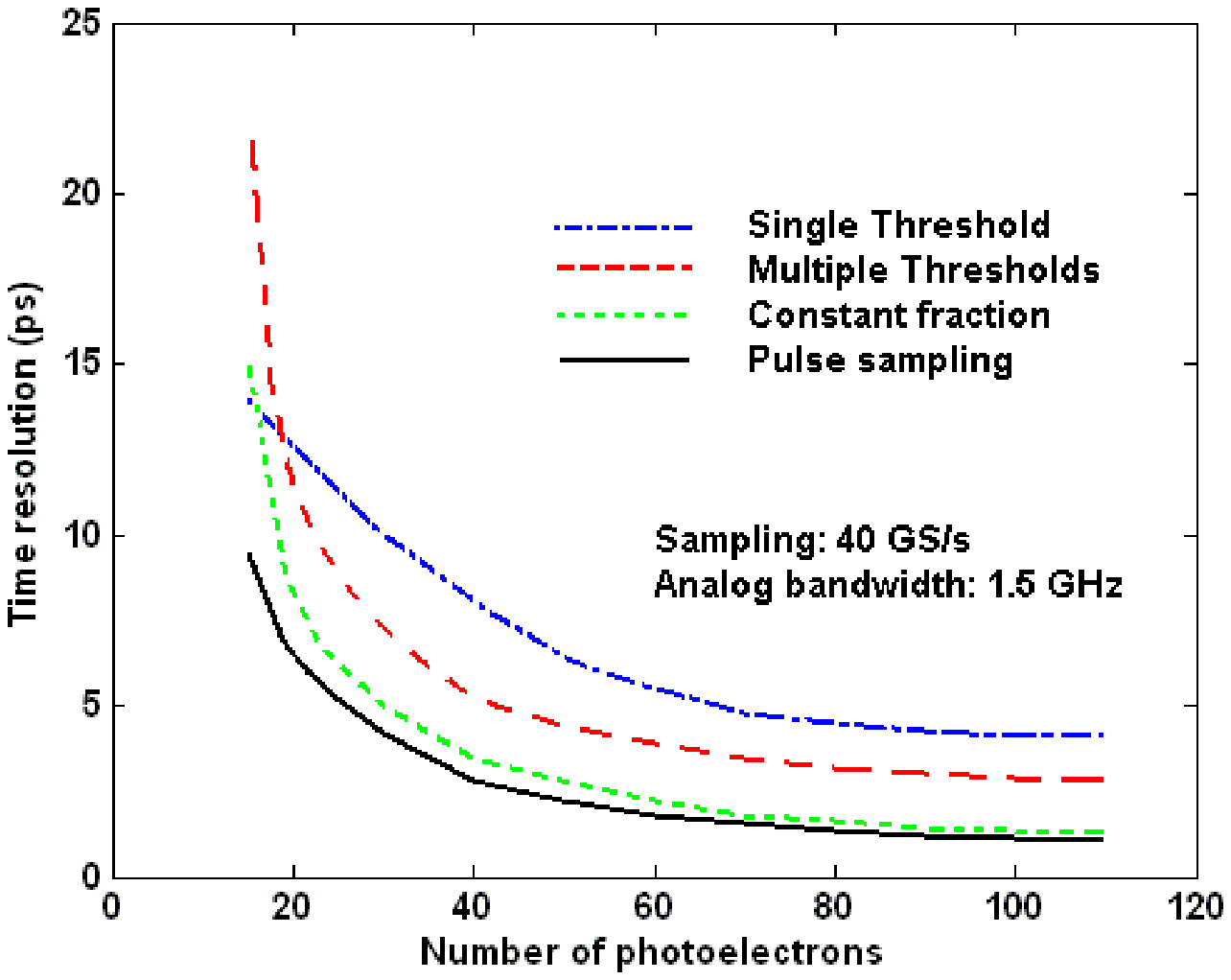}
\hfil
\includegraphics[angle=0,width=0.45\textwidth]{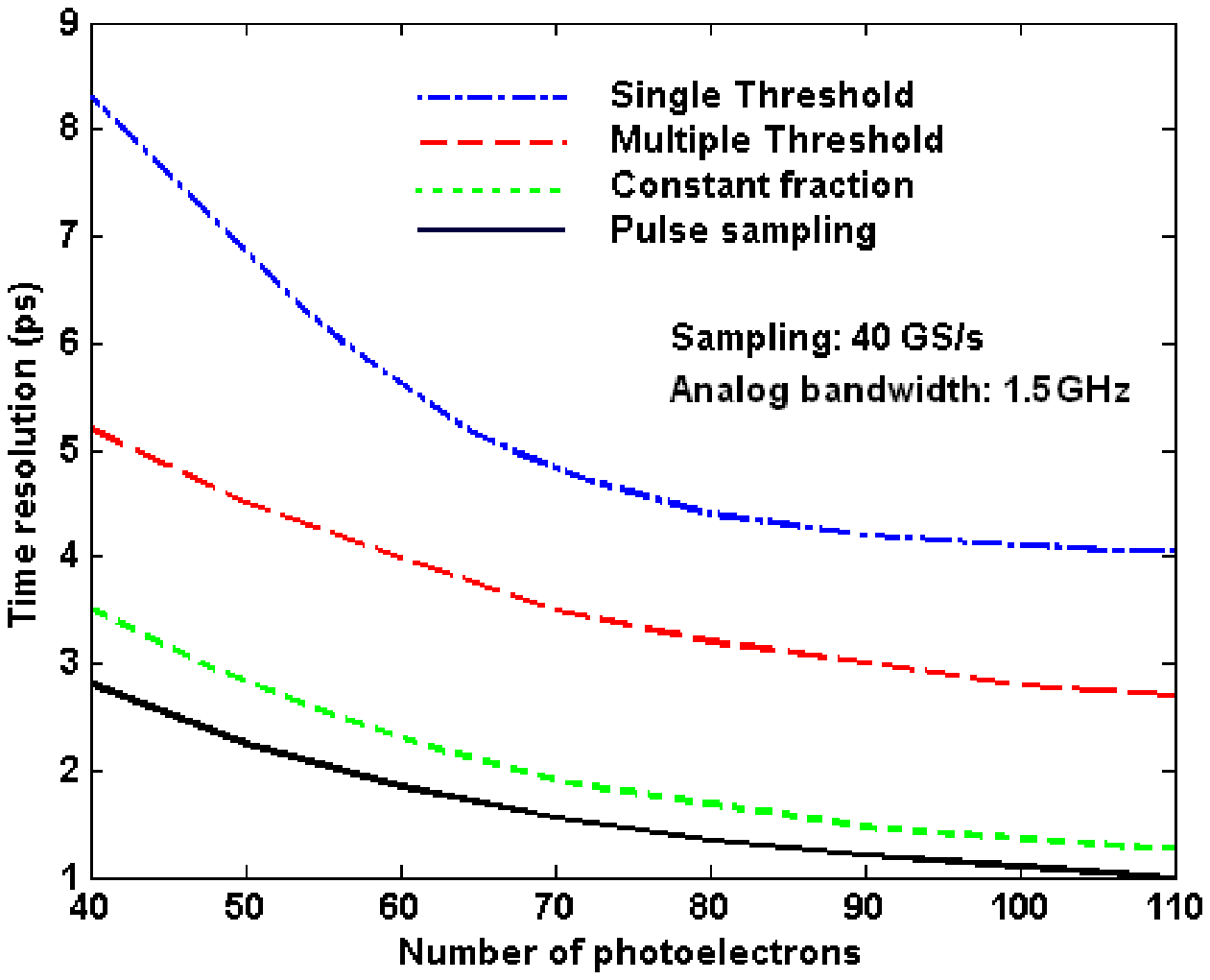}
\caption{Left:  Time resolution versus the number of primary photo-electrons, for the four different timing
techniques: one-threshold (blue, dashed-dots), constant fraction (green, dots), multiple threshold (red, dashed), and waveform sampling (black, solid) at 40 GSa/s. The analog bandwidth of the input to the sampling is taken to be 1.5 GHz, no sampling jitter
added. Right: The same plot with the abscissa expanded to cover from 40 to 110 photo-electrons, and the ordinate
expanded to cover 1 to 9 ps. } \label{fig:sampling_npe}
\end{figure}

 The simulation of pulse waveform sampling performs a number of
 samples of the signal voltage at equally spaced time intervals,
 digitized with a given precision in amplitude corresponding to the
 number of bits assumed in the A-to-D conversion, and adds a random
 jitter in the time of each sample.  Sampling rates between 10 and 60
 GSa/s have been simulated for A-to-D conversion precisions of between 4
 and 16-bits.  The analog bandwidth of the MCP device and associated
 front-end electronics are included in the simulations.

An iterative least-squares fit to a noiseless MCP template signal is
then applied to the data using the Cleland and Stern
algorithm that has been implemented for high resolution calorimetry
measurements with Liquid Argon~\cite{cleland_stern} .

\section{Results of the Simulations}
\label{results}
Figure 5 shows the time resolution versus the number of
photo-electrons for the four timing techniques.  The number of
photo-electrons is varied between between 15 and 110; the sampling
rate is 40 GSa/s, with no sampling jitter. The analog bandwidth is
assumed to be 1.5 GHz, and in the case of sampling, the digitization is taken to have a
precision of 16 bits.

The waveform sampling technique performs best of the four techniques, particularly for lower numbers of
photo-electrons. Note that the sampling technique is relatively insensitive to (random) clock jitter; 
at 40 GSa/s sampling; jitters smaller than 5 psec do not introduce significant degradation in the
resolution~\cite{jenkins,jitter}.

Table~\ref{tab:resolution} shows the timing resolution for each
technique for a set of parameters we have chosen as a base-line for
use in a future detector: an input signal of 50 photo-electrons, a
1.5 GHz analog bandwidth, and for each technique, a signal-to-noise
ratio of 80. All Monte-Carlo simulations were run with 300 events; the
corresponding statistical uncertainty is on the order of 5\%.

\begin{table}[htb]
\centering
\begin{tabular}{|l|c|}
\hline \hline Technique & Resolution \\
          & (ps) \\
\hline
Leading Edge       & 7.1\\
Multiple Threshold & 4.6\\
Constant Fraction  & 2.9\\
Sampling           & 2.3\\
\hline\hline
\end{tabular}
\vskip 0.1in
\caption{The predicted timing resolution for each
technique for an assumed input signal of 50 photo-electrons, a 1.5 GHz analog
bandwidth, and a signal-to-noise ratio of 80. No sampling jitter has
been added. The statistical uncertainty on each result is on the order of 5\%.}
\label{tab:resolution}
\end{table}

\begin{figure}[!bh]
\centering
\includegraphics[angle=0,width=0.45\textwidth]{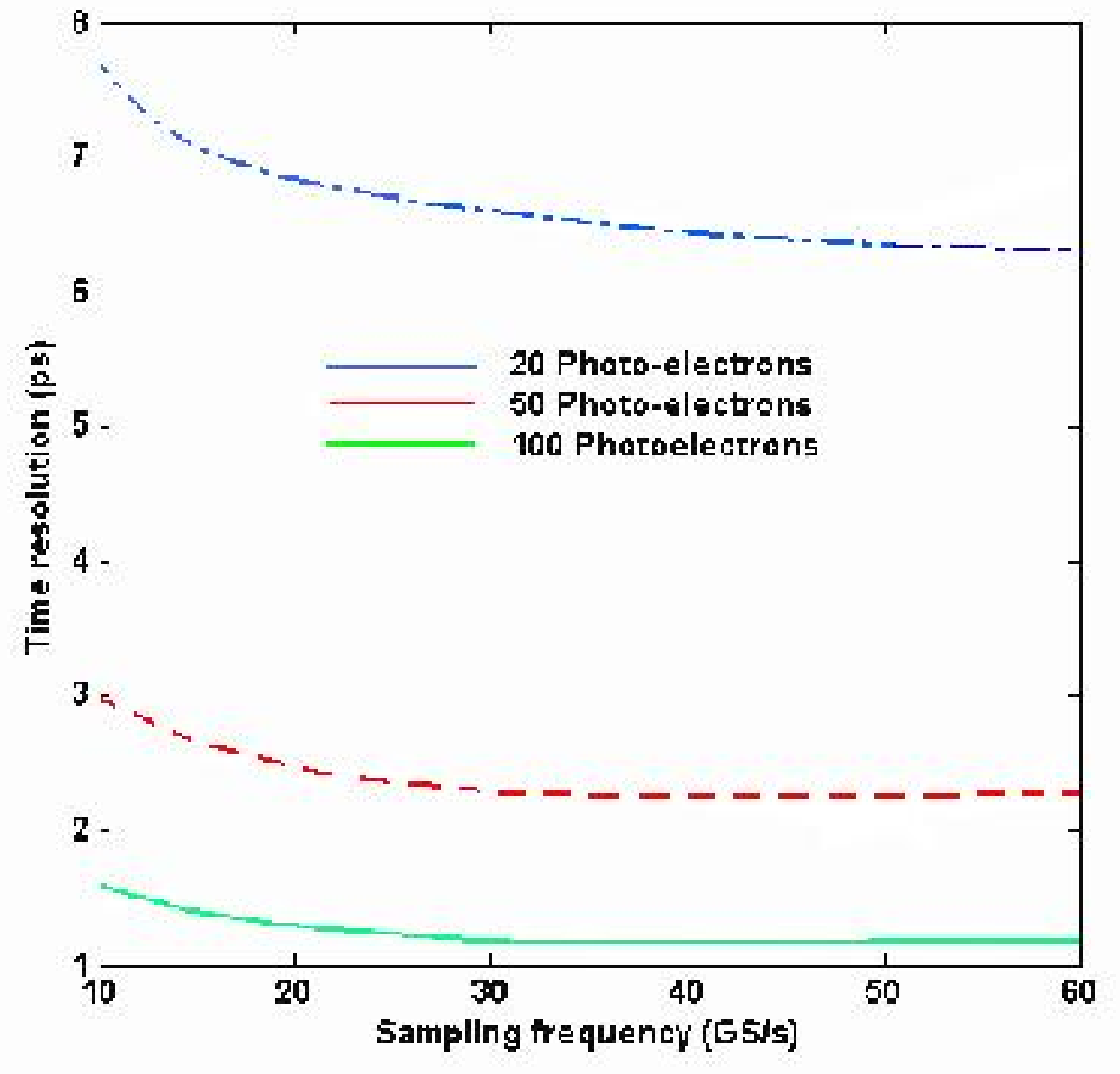}
\hfil
\includegraphics[angle=0,width=0.45\textwidth]{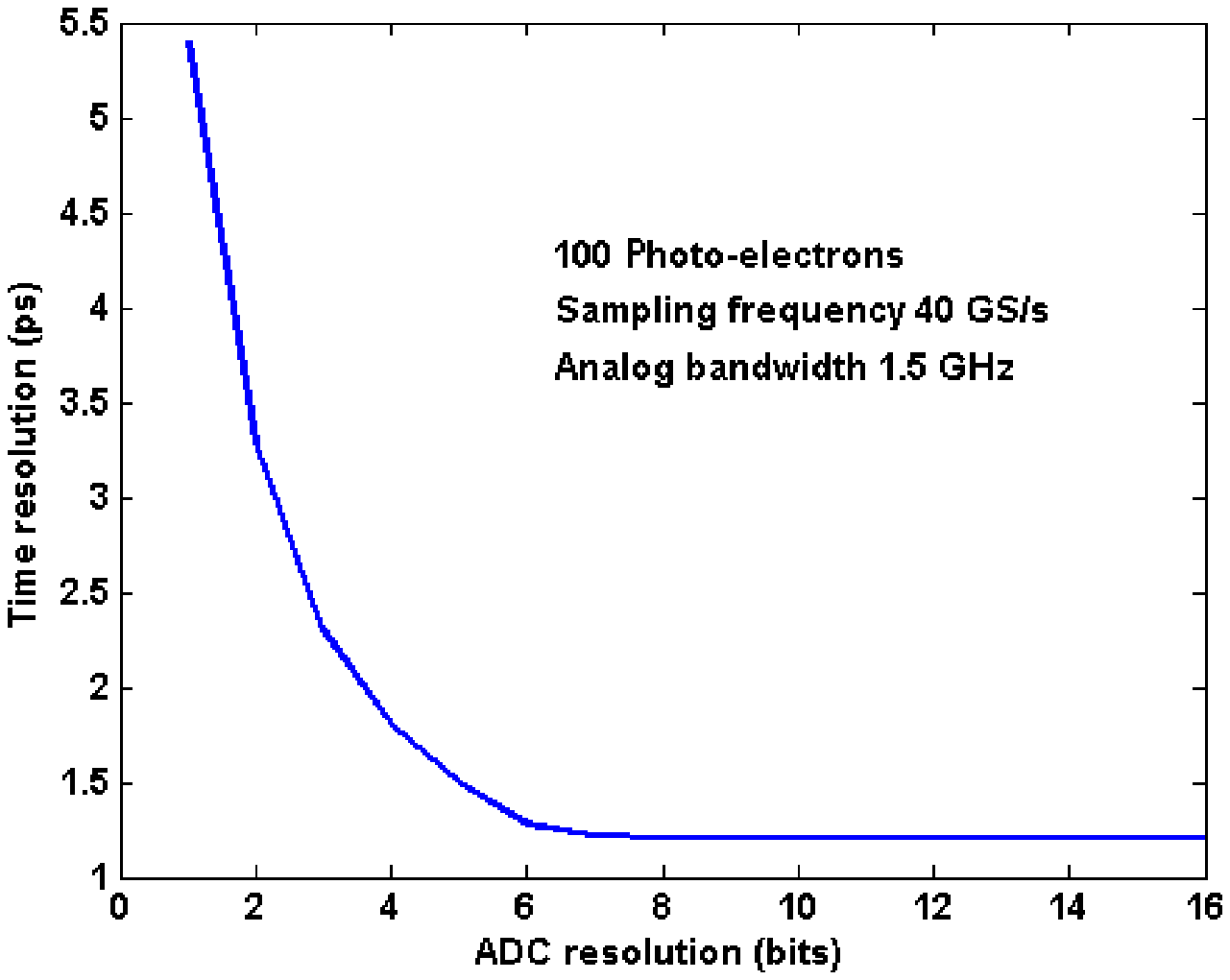}
\caption{ Pulse waveform sampling simulation results. Left: Time resolution versus sampling rate. The input
signals correspond to 20, 50 and 100 photo-electrons. The digitization precision is 16-bits, with no sampling
jitter added. The analog bandwidth is taken to be 1.5 GHz. Right: The timing resolution as a function of the
number of bits used in the digitization.  The sampling rate is 40 GSa/s, the analog bandwidth is 1.5 GHz. The signal
is taken to be 100 photo-electrons, with a signal-to-noise ratio of 164. No sampling jitter has been added. }
\label{fig:sampling_rate_nbits}
\end{figure}

As expected, the multiple threshold technique is a clear improvement compared to the one-threshold discriminator
for input signals above 10 photo-electrons.  The constant fraction discriminator also is a significant improvement
over a single-threshold for any input signal. However, the best method is waveform sampling, which achieves
resolutions below 3 picoseconds for input signals of 50 photo-electrons in the base-line case of a S/N ratio of
80, 1.5 GHz analog input bandwidth, and random sampling jitter less than 5 psec.

Figure~\ref{fig:sampling_rate_nbits} shows the sensitivity of sampling to
the digitization. The curve flattens out such that an
8-bit digitization is  sufficient to get a resolution of a few
picoseconds at a 40 GSa/s sampling rate. This greatly relaxes the
constraints on the analog- to-digital converter design (we note that
the A-to-D conversion does not have to be fast- the sampling is at 40
GSa/s, but the digitization rate is set by the occupancy requirements of
the application, and for the small pixel sizes we are considering for
most time-of-flight applications the digitization can be done in real
time at rates below 1 MHz).

\begin{figure}[!t]
\centering
\includegraphics[angle=0,width=0.45\textwidth]{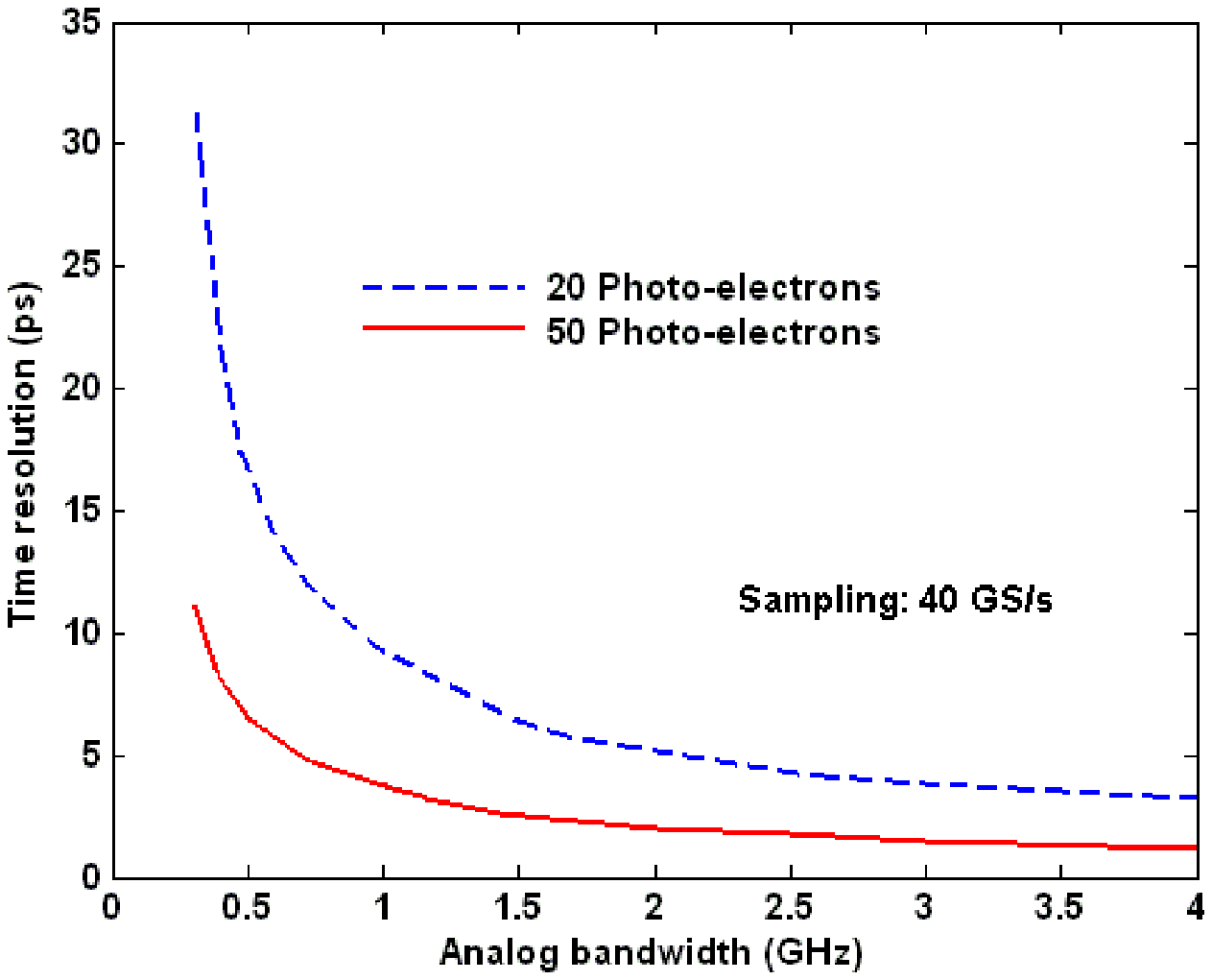}
\hfil
\includegraphics[angle=0,width=0.45\textwidth]{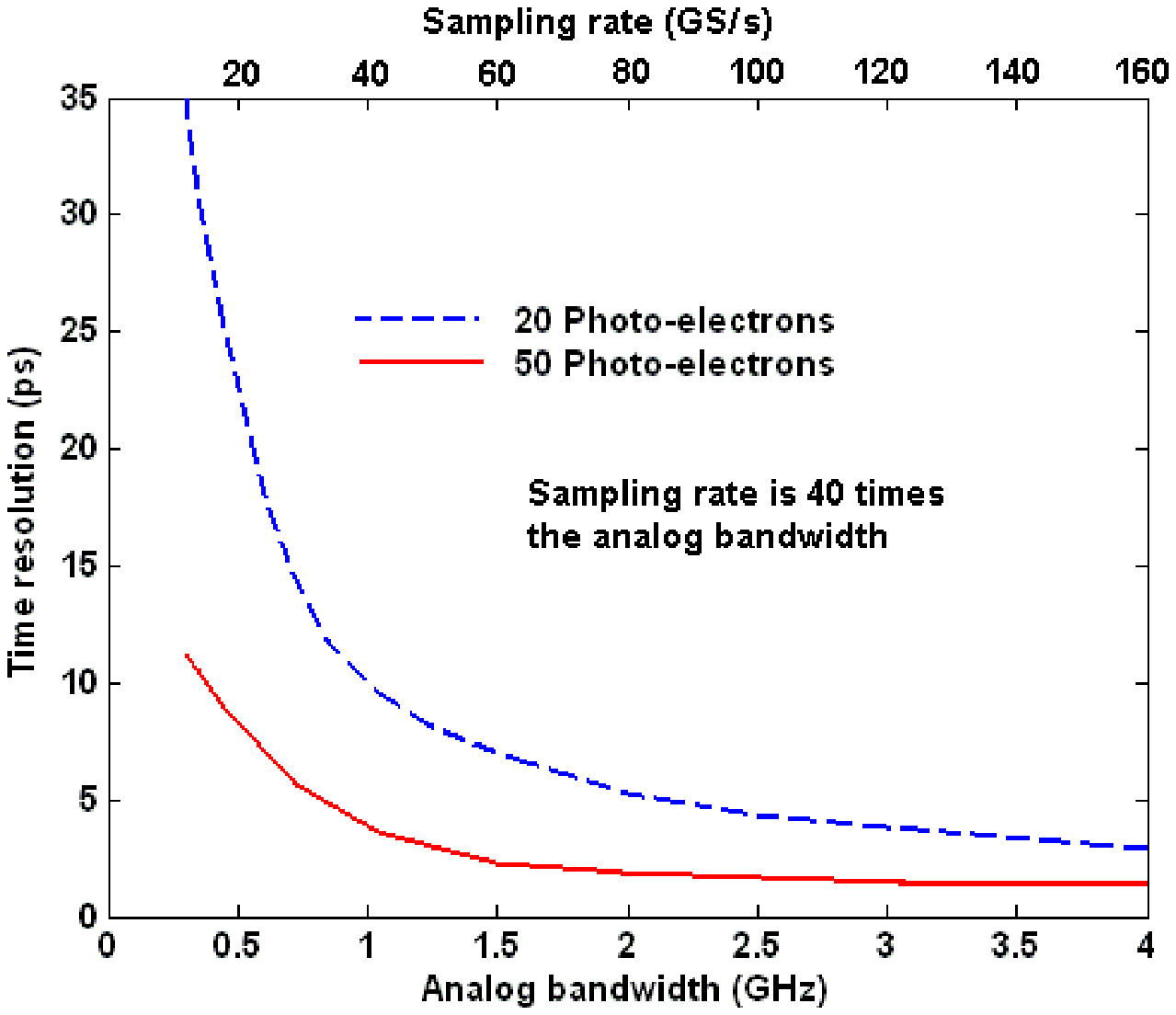}
\caption{ Pulse waveform sampling simulation results. Left: Time resolution versus analog bandwidth for a fixed
sampling rate of 40 GSa/s, for input signals of 20 and 50 photo-electrons and an 8-bit A-to-D precision. Right:
Time resolution versus analog bandwidth for a sampling rate proportional to the analog bandwidth, with 8-bit
digitization and a 2-ps timing jitter. } \label{fig:sampling_abw}
\end{figure}

We have simulated the dependence of the time resolution with the analog bandwidth for our baseline sampling rate
of 40 GSa/s, as shown in the left-hand plot of Figure~\ref{fig:sampling_abw}. The time resolution versus analog
bandwidth for a sampling rate proportional to the analog bandwidth is shown in the right-hand plot of
Figure~\ref{fig:sampling_abw}. In both cases, the time resolution improves with the analog bandwidth as expected,
but flattens above a 2 GHz analog bandwidth and a 80 GSa/s sampling rate, showing that there is not much to gain in
designing electronics beyond these limits.

\begin{figure}[!t]
\centering
\includegraphics[angle=0,width=0.65\textwidth]{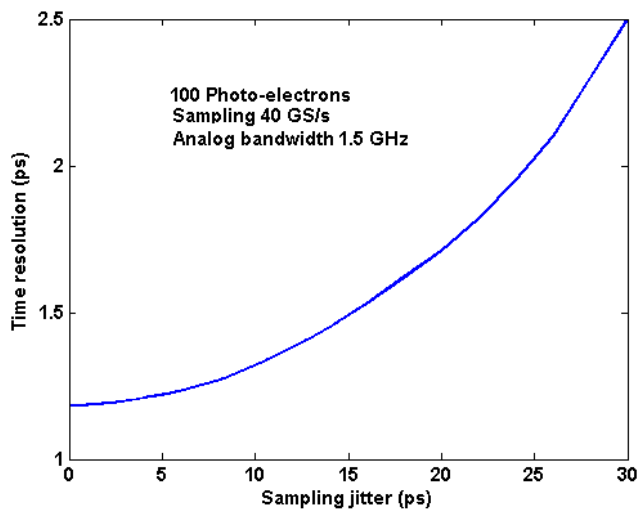}
\caption{Right: Time resolution vs sampling
jitter for input signals of 100 photo-electrons.  The sampling
rate is 40 GSa/s, the analog bandwidth is 1.5 GHz, and the digitization
is assumed to be with a 16-bit ADC.
}
\label{fig:sampling_jitter}
\end{figure}

The waveform sampling technique is also robust against random
sampling clock jitter provided there are enough samples on the leading
edge, as shown Figure~\ref{fig:sampling_jitter}.

\section{Summary and Conclusions}
\label{conclusions}

We have developed a simulation package based on MATLAB to model the time resolution for fast pulses from
photo-detectors.  Using the parameters measured from commercial micro-channel plate photo-multipliers, we have
simulated and compared the time-resolutions for four signal processing techniques: leading edge discriminators,
constant fraction discriminators, multiple-threshold discriminators and pulse waveform sampling.  We find that
timing using pulse waveform sampling gives the best resolution in many cases, particularly in the presence of
white noise and a substantial signal, such as fifty photo-electrons. With micro-channel plate photo-detectors, our
simulations predict that it should be possible to reach a precision of several picoseconds or better with pulse
waveform sampling given large-enough input signals.  At high sampling rates of the order of 40 GSa/s, a relatively
low precision digitization (8-bit) can be used.

\section{Acknowledgments}
\label{acknowledgements}

We are indebted to Dominique Breton, Eric Delagnes, Stefan Ritt, and 
their groups working on fast waveform sampling. We
thank John Anderson, Karen Byrum, Chin-Tu Chen, Gary Drake, Camden Ertley, Daniel
Herbst, Andrew Kobach, Jon Howorth, Keith Jenkins, Chien-Min Kao, 
Patrick Le Du, 
Tyler Natoli, Rich Northrop, Erik Ramberg, Anatoly Ronzhin, Larry Ruckman, Andrew Wong, David Salek,
Greg Sellberg, Scott Wilbur, and Jerry Va'Vra for valuable
contributions.  We thank Paul Hink and Paul Mitchell of
Burle/Photonis for much help with the MCP's, and Larry Haga of
Tektronix Corporation for his help in acquiring the Tektronix TDS6154C
40 GSa/s sampling oscilloscope on which much of this work is based.
This work was supported in part by the University of Chicago, the
National Science Foundation under grant number 5-43270, and the
U.S. Department of Energy Advanced Detector Research program.


\begin{thebibliography}{99}

\bibitem{CDF}See, for example, D. Acosta et al. (CDF Collaboration);
{\it The Performance of the CDF Luminosity Monitor.}
Nucl. Instr. Meth. {\bf A518} (2004) 605-608.

\bibitem{TOF_review} W. Klempt.
{\it  Review of Particle Identification by Time of Flight techniques.}
Nucl. Instr. Meth. {\bf A433} (1999) 542-553.

\bibitem{Wiza} J.L. Wiza.
{\it Micro-channel Plate Detectors.}
Nucl. Instr. Meth. {\bf 162} (1979) 587-601.

\bibitem{Credo_Rome} T. Credo, H. Frisch, H. Sanders, R. Schroll, F. Tang.
{\it Picosecond Time of Flight for Particle Identification at High Energy Physics Colliders.}
Proceedings of the Nuclear Science Symposium, Rome (2004), 586.


\bibitem{Inami} K. Inami, N. Kishimoto, Y. Enari, M. Nagamine, and T. Ohshima.
{\it Timing properties of MCP-PMT.}
Nucl. Instr. Meth. {\bf A560} (2006) 303-308.

\bibitem{Vavra} J. Va'vra, J. Benitez, J. Coleman, D. W. G. Leith,
G. Mazaher, B. Ratcliff and J. Schwiening.
{\it A 30 ps Timing Resolution for Single Photons with Multi-pixel Burle MCP-PMT.}
Nucl. Instr. Meth. {\bf A572} (2007) 459-462.

\bibitem{LHC_photon} At a high luminosity machine such as the LHC
there are many collisions per beam crossing, making associating
photons from a Higgs decay with a specific vertex difficult, to pick
one example. This application would require the conversion of the
photon and a simultaneous precision measurement of the time and
position.

\bibitem{gunion} See, for example, C.H. Chen and J. F. Gunion.
{\it Probing Gauge-Mediated Supersymmetry Breaking Models at the Tevatron via Delayed Decays of the Lightest Neutralino.}
Phys. Rev. D 58 075005 (1998).

\bibitem{toback_delayed_photon} T. Aaltonen et al (CDF Collaboration).
ArXiv: 0804.1043 [hep-ex], submitted to Phys. Rev. D.

\bibitem{LHC_forward} See, for example,\\
http://hepwww.rl.ac.uk/accel/forum/2007/Cosenerswattsapr07.pdf


\bibitem{MANX} The MANX Project Proposal
https://mctf.fnal.gov/meetings/2007-1/04\_05/project-narrative-06er86282-6dmanx-v2-w-appendices.pdf/view

\bibitem{nicholson} Howard Nicholson, private communication.


\bibitem{porat} D.I. Porat
{\it Review Of Sub-nanosecond Time Interval Measurements};
IEEE Trans. Nucl. Sci. {\bf 20} (1973) 36.


\bibitem{genat}  J.F. Genat.
{\it High Resolution Time to Digital Converters.}
Nucl. Inst. Meth. {\bf A315} (1992) 411-414.


\bibitem{kalisz} J. Kalisz 
{\it Review of Methods for Time Interval Measurements with Picosecond Resolution}, Institute of Physics
 Publishing, Metrologia, 41 (2004) 17-32;
 http://www.iop.org/EJ/abstract/0026-1394/41/1/004


\bibitem{mantyniemi} An extensive list of references on timing measurements can be found in:  
A. Mantyniemi, MS Thesis, Univ. of Oulu, 2004; ISBN 951-42-7460-I; ISBN 951-42-7460-X;
http://herkules.oulu.fi/isbn951427461X/isbn951427461X.pdf


\bibitem{cova} S. Cova et al.  
{\it Constant Fraction Circuits for Picosecond Photon Timing with Micro-channel Plate Photomultipliers.}
Review of Scientific Instruments, 64-1 (1993) 118-124.

\bibitem{heejong} H. Kim et al.  
{\it Electronics Developments for Fast Timing PET Detectors.}  
Symposium on Radiation and Measurements
Applications. June 2-5 (2008), Berkeley CA, USA.

\bibitem{breton} D. Breton, E. Auge, E. Delagnes, J. Parsons, W
Sippach, V. Tocut.
{\it The HAMAC rad-hard Switched Capacitor Array.}
ATLAS note. (2001).

\bibitem{delagnes} E. Delagnes, Y. Degerli, P. Goret, P. Nayman,
F. Toussenel, and P. Vincent.  
{\it SAM : A new GHz sampling ASIC for the HESS-II Front-End.} 
Cerenkov Workshop (2005)

\bibitem{ritt} S. Ritt.  
{\it Design and Performance of the 5 GHz Waveform
Digitizer Chip DRS3.} 
Submitted to Nuclear Instruments and Methods, (2007).

\bibitem{varner} G. Varner, L.L. Rudman, A. Wong.
{\it The First version Buffered Large Analog Bandwidth (BLAB1) ASIC for high Luminosity Colliders and Extensive Radio Neutrino Detectors.} 
Nucl. Inst. Meth. {\bf A591} (2008) 534.

\bibitem{photek} J. Milnes and J. Howorth,  
Photek Ltd; Proc. SPIE 5580 (2005) 730-740.

\bibitem{inami} K. Inami.
{\it Timing properties of MCP-PMTs. Proceedings of Science.}
International Workshop on new Photon-Detectors, 
June 27-29 (2007). Kobe University, Japan.

\bibitem{bondarenko} G.Bondarenko, B. Dolgoshein et al.
{\it Limited Geiger Mode Silicon Photodiodes with very high Gain.}  
Nuclear Physics B, 61B (1998) 347-352.


\bibitem{shannon} C.E. Shannon.
{\it A Mathematical Theory of Communication.}
The Bell System Technical Journal, 27 (1948)
379-423 623-656.

\bibitem{matlab} MathWorks, 3 Apple Hill Drive, Natick, MA, USA.
 The MATLAB source is available from the authors.

\bibitem{definitions} The signal-to-noise ratio is defined as the
  ratio of the maximum of the amplitude of the pulse to the rms
  amplitude of the noise. The analog-bandwidth is defined as the
  frequency at which the system response has dropped by 3-db.

\bibitem{Photonis} Photonis/Burle Industries, 1000 New Holland Av,
Lancaster PA, 17601.

\bibitem{Tektronix} Tektronix INC, PO Box 500, Beaverton, OR 97977.

\bibitem{ANL_laser} C. Ertley, J. Anderson, K. Byrum, G. Drake, E. May.
http://www.hep.anl.gov/ertley/windex.html


\bibitem{cleland_stern}   W.E. Cleland and E.G. Stern.
{\it Signal Processing considerations for Liquid Ionization Calorimeters in a High Rate Environment.}
Nucl. Instr. Meth. {\bf A338} (1994) 467-497.

\bibitem{jenkins} K.A. Jenkins, A.P. Jose, D.F Heidel
{\it An On-chip Jitter Measurement Circuit with Sub-picosecond Resolution.}
Proceedings of the 31st European Solid State Circuits Conference, 12 (2005) 157-160.

\bibitem{jitter} Systematic errors in sampling may be calibrated out
with the use of extra calibration channels in the front-end readout.



\end{thebibliography}
\end{document}